\documentclass[]{proarticle}
\usepackage{multicol}
\usepackage{graphicx}
\usepackage{amssymb}
\usepackage{epsfig}

\def\bea{\begin{eqnarray}}
\def\eea{\end{eqnarray}}
\begin{document}
\title{$U(N)$ spinning particles and higher spin fields on K\"ahler backgrounds}
\author{Roberto Bonezzi}
\maketitle
\address{Dipartimento di Fisica, Universit\`{a} di Bologna and INFN, sezione di Bologna, via Irnerio 46, I-40126 Bologna, Italy}
\eads{bonezzi@bo.infn.it}
\begin{abstract}
In this short contribution we will review the quantization of $U(N)$ spinning particles with complex target spaces, producing equations for higher spin fields on complex backgrounds. We will focus first on flat complex space, and subsequently discuss how to extend our model on suitable K\"ahler manifolds. In the final section, we will specialize to $(p,q)$-forms on arbitrary K\"ahler spaces and present their one-loop effective actions as well as issues related to Hodge duality.
\end{abstract}


\begin{multicols}{2}
\section{Introduction}

Spinning particle models \cite{Gershun:1979fb,Howe:1988ft,Kuzenko:1995mg,Bastianelli:2007pv} have been a useful framework to study fields of different spins in a first-quantized approach.
Wordline techniques can indeed give manageable representations of one-loop quantities in quantum field theories, such as effective actions, amplitudes and anomalies (see for a review \cite{Schubert:2001he,Bastianelli:2006rx} and references therein), and allow to describe ordinary higher spin fields in first quantization . We will present here a class of spinning particles enjoying a $U(N)$-extended supersymmetry on the worldline \cite{Marcus:1994mm,Bastianelli:2009vj}, that naturally live on complex manifolds and give rise to complex higher spin equations.
Although a direct spacetime interpretation of these models is prevented by the complex nature of the target space, they are useful playgrounds to study issues related to the quantization of higher spins. They can as well provide insights in the study of K\"ahler geometries, supersymmetric field theories and QFT's in curved backgrounds.

More precisely, spinning particles are quantum mechanical models, enjoying local supersymmetries on the worldline. For instance, particles with $N$ local real
supersymmetries describe spin $N/2$ fields in four dimensional spacetime. The constraints on the Hilbert space of the particle theory, arising from the gauging, translate into a set of differential equations for the spacetime field, viewed as the wave function of the quantum mechanical model.
Interactions with scalars, gauge fields and gravity can be achieved, when allowed, by coupling the particle theory to suitable backgrounds. Quantizing the particle, one can recover in a rather simple way various useful objects of the related QFT, such as effective actions, n-point correlation functions, anomalies and
so on. The path integral quantization for these non-linear sigma models, arising in gravitational backgrounds, requires regularization \cite{Bastianelli:2006rx,Bonezzi:2008gs,Bastianelli:2011cc}. This is essentially related to ill-defined products of distributions in perturbative computations, and we will use Time Slicing regularization (TS) in all the computations that will be showed.

We organize the paper as follows: in the next section we will present the $U(N)$ spinning particle and its Dirac quantization in flat complex space. The end of the section will be devoted to the coupling to a curved K\"ahler background. In the last section we will focus on the $U(2)$ model, studied in \cite{Bastianelli:2012nh}, describing $(p,q)$-forms on an arbitrary K\"ahler space. The local proper time expansion of the one-loop effective action will be briefly sketched, along with the first Seeley-DeWitt coefficients. We will comment at the end on exact relations between Hodge dual forms that have been extracted from the particle model.
%
\section{The $U(N)$ spinning particle}\label{sec2}
%
The graded phase space of the model is spanned by complex coordinates of $\mathbb{C}^d$ and momenta: $x^\mu(t)$, $\bar x^{\bar\mu}(t)$ and $p_\mu(t)$, $\bar p_{\bar\mu}(t)$, along with the fermionic superpartners $\psi^\mu_i(t)$ and $\bar\psi_\mu^i(t)$, with $i=1,2,...,N$. They obey equal time canonical (anti)-commutation relations:
\begin{equation}\label{canonical commutators}
[x^\mu,p_\nu]=i\,\delta^\mu_\nu\;,\quad[\bar x^{\bar\mu},\bar p_{\bar\nu}]=i\,\delta^{\bar\mu}_{\bar\nu}\;,\quad \{\psi^\mu_i,\bar\psi_\nu^j\}=\delta^\mu_\nu\delta^j_i\;.
\end{equation}
Quadratic operators constructed from the basic variables provide generators for the $U(N)$-extended worldline SUSY:
\begin{eqnarray}\label{UN generators flat}\nonumber
H &=& p_\mu\bar p^\mu\;,\quad Q_i=\psi^\mu_i\,p_\mu\;,\quad\bar Q^i=\bar\psi^{\bar\mu i}\,\bar p_{\bar\mu}\;,\\
J^i_j &=& \frac12[\psi^\mu_j,\bar\psi^i_\mu]=\psi^\mu_j\bar\psi_\mu^i-\frac d2\,\delta^i_j\;,
\end{eqnarray}
where we raise and lower indices by means of the flat complex metric $\delta_{\mu\bar\nu}$ and its inverse. $H$, $Q_i$ and $\bar Q^i$ are the hamiltonian and supercharges, respectively generating worldline translations and supersymmetries, while $J^i_j$ generates $U(N)$ $R$-symmetry rotations. Given the (anti)-commutation relations (\ref{canonical commutators}), the above generators obey an extended supersymmetry algebra with $R$-symmetry group $U(N)$:
\begin{eqnarray}\label{UN superalgebra}\nonumber
\left[J^i_j, Q_k\right] &=& \delta^i_k\,Q_j\;,\quad \left[J^i_j, \bar Q^k\right]=-\delta^k_j\,\bar Q^i\;,\\
\left[J^i_j,J^k_l\right]&=& \delta^i_l\,J^k_j-\delta^k_j\,J^i_l\;,\quad\left\{Q_i,\bar Q^j\right\}=\delta^j_i\,H
\end{eqnarray}
the other (anti)-commutators being zero. In order to construct a worldline action that is invariant under \emph{local} symmetries generated by (\ref{UN generators flat}), we couple the generators to one-dimensional gauge fields: an einbein $e(t)$ for local worldline translations, complex gravitinos $\chi_i(t)$ and $\bar\chi^i(t)$ for local supersymmetries and a one-dimensional $U(N)$ gauge field $a^i_j(t)$. Together with the symplectic kinetic terms, we obtain the phase space action for the $U(N)$ spinning particle:
\begin{eqnarray}\nonumber
S=\int_0^1dt\!\!\!\!\!\!\!\!\!\!\!&&\left[p_\mu\dot x^\mu+\bar p_{\bar\mu}\dot{\bar x}^{\bar\mu}+i\bar\psi_\mu^i\dot\psi_i^\mu-eH-i\bar\chi^iQ_i\right.\\
&&\left.-i\chi_i\bar Q^i-a^i_j\left(J^j_i-s\delta^j_i\right)\right]\;,\label{phase space action UN spinning}
\end{eqnarray}
where we added a Chern-Simons coupling $s$ that is quantized as $s=m-\frac d2$ for integer $m$.
The equations of motion for the one-dimensional gauge fields constrain the classical generators to vanish. At the quantum level, the operators (\ref{UN generators flat}) impose constraints on the Hilbert space, by requiring that they annihilate physical states:
\begin{equation}\nonumber
\left|\Phi\right\rangle\in \mathcal{H}_{\rm phys}\quad\Leftrightarrow\quad T_A\left|\Phi\right\rangle=0\;,
\end{equation}
where $T_A=(H,Q_i,\bar Q^i,J^i_j-s\delta^i_j)$. This is allowed since (\ref{UN superalgebra}) is a first class superalgebra, and amounts to Dirac quantization. The above constraints will be the aforementioned higher spin equations obeyed by spacetime fields, that sit as $x$-dependent coefficients inside $\left|\Phi\right\rangle$, when expanded in terms of position eigenstates $\left|x\right\rangle$.

\subsection{Dirac quantization in flat space}

In a Schroedinger picture, we shall realize the fermionic oscillator algebra in (\ref{canonical commutators}) treating the $\psi^\mu_i$ operators as Grassmann-odd variables, and the $\bar\psi^i_\mu$ as odd derivatives thereof: $\bar\psi^i_\mu\sim\frac{\partial}{\partial\psi_i^\mu}$.
The states in the Hilbert space will have a finite Taylor expansion in powers of $\psi$'s, so that the coefficients of the expansion are multi-form spacetime fields with only holomorphic indices. The $J^i_j-s\delta^i_j$ constraints are purely algebraic on the multi-forms, and impose irreducibility conditions with respect to $U(d)$. After imposing them, the only states that survive in the Hilbert space correspond to tensors with $N$ blocks of $m$ antisymmetric indices, with $s=m-\frac d2$:
$$
\left|F\right\rangle\sim F_{\mu^1[m],...,\mu^N[m]}(x,\bar x)\;,
$$
where we denoted $\mu[k]:=[\mu_1...\mu_k]$.
Symmetry between block exchanges ensures they belong to a $U(d)$ rectangular Young tableaux with $m$ rows and $N$ columns. On the spacetime fields
the supercharges $Q_i$ and $\bar Q^i$ act as Dolbeault operators\footnote{On a complex (p,q)-form $A_{\mu_1...\mu_p\bar\nu_1...\bar\nu_q}dz^{\mu_1}\wedge...dz^{\mu_p}\wedge d\bar z^{\bar\nu_1}...\wedge d\bar z^{\bar\nu_q}$ the Dolbeault operator acts as an holomorphic exterior derivative: $\partial:=dz^\mu\partial_\mu$} and their hermitian conjugates, generalized to multi-forms such that $Q_k\sim-i\partial_{(k)}$ meaning that it antisymmetrizes the derivative only among the indices of the $k$-th block.
The remaining constraints have then the form of generalized Bianchi and Maxwell equations, \emph{i.e.}
\begin{eqnarray}\nonumber
&&Q_i\left|F\right\rangle=0 \;\sim\; \partial_{[\mu}F_{\mu_1...\mu_m],...,\nu_1...\nu_m}=0\;, \\
&&\bar Q^i\left|F\right\rangle=0\;\sim\; \bar\partial^\mu F_{\mu\mu_2...\mu_m,...,\nu_1...\nu_m}=0\;.
\label{HS equation curvature}
\end{eqnarray}

It is natural to interpret the $F$ fields as higher spin curvatures that obey Maxwell-like equations. It is indeed possible to solve the first of (\ref{HS equation curvature}) by introducing a gauge potential $\phi$ as $\left|F\right\rangle=Q_1Q_2...Q_N\left|\phi\right\rangle$ or, in tensor language:
\begin{equation}\label{potential}
F_{\mu^1[m],...,\mu^N[m]}=\partial_{\mu^1}...\partial_{\mu^N}\phi_{\mu^1[m-1],...,\mu^N[m-1]}\;,
\end{equation}
where each set of indices $\mu^k$ is antisymmetrized: $\mu^k\mu^k[m-1]:=[\mu^k_1...\mu^k_m]$.

Let us notice that for $N>1$ the Maxwell equation (\ref{HS equation curvature}) on the potential is higher derivative. We also mention that for $N=1$ we have ordinary $(p,0)$-forms $\phi_{(p,0)}$, carefully studied in \cite{Bastianelli:2011pe}, while for $m=2$ the gauge potentials are completely symmetric tensors $\phi_{\mu_1...\mu_N}$. Since $(Q_i)^2=0$ for each $i$, it is easy to see that the curvature $F$ is indeed invariant under a gauge transformation of the form $\delta\left|\phi\right\rangle=Q_i\left|\Lambda^i\right\rangle$, that is
\begin{eqnarray}\nonumber
\delta\phi_{\mu^1[p],...,\mu^N[p]}=&&\!\!\!\!\!\!\!\!\partial_{\mu^1}\Lambda^{(1)}_{\mu^1[p-1],...,\mu^N[p]}\\
&&\!\!\!\!\!\!\!\!+\ldots+\partial_{\mu^N}\Lambda^{(N)}_{\mu^1[p],...,\mu^N[p-1]}
\label{gauge trans}
\end{eqnarray}
We already noticed that the Maxwell-like equations are higher derivative in terms of the gauge field: $\bar Q^iQ_1...Q_N\left|\phi\right\rangle=0$. It is actually possible to introduce a second order wave operator, analogous to the Fronsdal-Labastida one, and reduce the field equations to second order: $\left(-H+Q_i\bar Q^i\right)\left|\phi\right\rangle=Q_i Q_j\left|\rho^{ij}\right\rangle$ by means of an auxiliary compensator $\left|\rho^{ij}\right\rangle$.
In tensor language it reads
\begin{eqnarray}\nonumber
&&\!\!\!\!\!\!\!\!\!\!\!\!\partial_\nu\bar\partial^\nu\phi_{\mu^1[p],...,\mu^N[p]}-\sum_{i=1}^N\partial_{\mu^i}\bar\partial^\nu\phi_{\mu^1[p],..,\nu\mu^i[p-1],...,\mu^N[p]}=\\
&&=\sum_{i\neq j}\partial_{\mu^i}\partial_{\mu^j}\rho^{(ij)}_{\mu^1[p],..,\mu^i[p-1],..,\mu^j[p-1],..,\mu^N[p]}\;.
\label{HS equation potential}
\end{eqnarray}
The field equations are invariant under the gauge transformations (\ref{gauge trans}), provided that the compensator field transforms as the divergence of the gauge parameter. It turns out that it is possible to gauge fix the compensators to zero, at the price of having transverse gauge parameters\footnote{For related issues on real spacetime, see for instance \cite{Campoleoni:2012th,Francia:2012rg}}: $\bar\partial\cdot\Lambda=0$.

\subsection{Coupling to curved space}

We analyze here the changes needed to couple the spinning particle to an arbitrary background metric. Let us consider as target space a $D=2d$ dimensional K\"ahler manifold, equipped with a metric $g_{\mu\bar\nu}(x,\bar x)$ in holomorphic coordinates. Having in mind the minimal coupling, it is sufficient to replace suitably covariantized constraints in the action (\ref{phase space action UN spinning}). To this aim, we define $U(d)$ ``Lorentz'' generators $M^\mu_\nu=\frac12[\psi^\mu_i,\bar\psi_\nu^i]$ so that we can construct covariant momenta and supercharges\footnote{The $g$ factors ensure that $(Q_i)^\dagger=\bar Q^i$}:
\begin{eqnarray}
&&\!\!\!\!\!\!\!\pi_\mu=p_\mu+i\,\Gamma^\lambda_{\mu\nu}\,M^\nu_\lambda\;,\quad\bar\pi_{\bar\mu}=\bar p_{\bar\mu}\;,\\
&&\!\!\!\!\!\!\!Q_i=\psi^\mu_i\,g^{\frac12}\pi_\mu g^{-\frac12}\;,\quad \bar Q^i=\bar\psi_\mu^i\,g^{\mu\bar\nu}\,g^{\frac12}\bar\pi_{\bar\nu} g^{-\frac12}
\end{eqnarray}
The superalgebra (\ref{UN superalgebra}) is deformed by the target space geometry, namely one has
\begin{equation}\label{QQbar curved}
\left\{Q_i,\bar Q^j\right\}=\delta^j_i\,H_0-\psi_i^\mu\bar\psi^{j\bar\nu}\,R_{\mu\bar\nu\lambda\bar\sigma}\,M^{\lambda\bar\sigma}\;,
\end{equation}
where the minimally covariantized hamiltonian reads $H_0=g^{\frac12}g^{\mu\bar\nu}\bar\pi_{\bar\nu}\pi_\mu g^{-\frac12}$. One can see from (\ref{QQbar curved}) that the algebra is no longer first class, and hence the model is inconsistent, on general K\"ahler backgrounds. Important exceptions are the cases $N=1,2$, presented in the next section, that can be quantized on any curved background, and represent differential forms. For $N>2$ one can still quantize the model on particular backgrounds. For instance, it is possible to quantize the spinning particle for general $N$, that is for arbitrary ``spin'', on K\"ahler spaces with constant holomorphic curvature \cite{Bastianelli:2009vj}, \emph{i.e.}
$$
R_{\mu\bar\nu\lambda\bar\sigma}=\Lambda\,\left(g_{\mu\bar\nu}g_{\lambda\bar\sigma}+g_{\mu\bar\sigma}g_{\lambda\bar\nu}\right)\;.
$$

\section{N=2, (p,q)-forms on K\"ahler spaces}

We shall focus in this section to the model with $N=2$ on an arbitrary K\"ahler manifold with metric $g_{\mu\bar\nu}$. We decide to realize the fermionic operators in a slightly different way: here we will treat $\psi^\mu_1$ and $\bar\psi^{2\bar\mu}$ as odd coordinates, and $\bar\psi^1_\mu$, $\psi_{2\bar\mu}$ as derivatives thereof. In this way, states in the Hilbert space are $(p,q)$-forms. In \cite{Bastianelli:2012nh}, several theories of differential forms were investigated by using the $U(2)$ spinning particle. Here we restrict ourselves to present one of those models. To obtain the model we are interested in, it is sufficient to gauge only the $U(1)\times U(1)$ subgroup of the $R$-symmetry $U(2)$ generated by $J^1_1$ and $J_2^2$. In such a case one is free to have two different Chern-Simons couplings $s_1$ and $s_2$, compared to $s$ in (\ref{phase space action UN spinning}). This allows to fix independently the eigenvalues of the $J^1_1$ and $J^2_2$ constraints, and physical states will be forms $F_{(m,n)}$, for given arbitrary $m,n$. The four supercharges are realized now as Dolbeault operators and their hermitian conjugates: $\partial$, $\bar\partial$ and $\partial^\dagger$, $\bar\partial^\dagger$, while the superalgebra (\ref{UN superalgebra}) closes on the hamiltonian that acts as the Hodge laplacian
$$
\triangle=-\{\partial,\partial^\dagger\}=\frac{\nabla^2}{2}+\frac12\,R_{\mu\bar\nu\lambda\bar\sigma}\,M^{\mu\bar\nu}M^{\lambda\bar\sigma}\;.
$$
Bianchi equations can be locally integrated by introducing a potential: $F_{(p+1,q+1)}=\partial\bar\partial A_{(p,q)}$. In this particular model, even if we are dealing with differential forms, Maxwell equations ($\partial^\dagger F=\bar\partial^\dagger F=0$) are higher derivative with respect to the potential $A$. As we did in the general case, it is possible to have second order field equations by introducing a compensator:
\begin{equation}\label{field eq N=2}
\left(\triangle+\partial\partial^\dagger+\bar\partial\bar\partial^\dagger\right)A_{(p,q)}=\partial\bar\partial\rho_{(p-1,q-1)}\;.
\end{equation}
The equations (\ref{field eq N=2}) are gauge invariant under the combined transformations of the gauge field and compensator
\begin{eqnarray}\nonumber
\delta A_{(p,q)}&=&\partial\Lambda^1_{(p-1,q)}+\bar\partial\Lambda^2_{(p,q-1)}\;,\\
\delta\rho_{(p-1,q-1)}&=&\bar\partial^\dagger\Lambda^1_{(p-1,q)}-\partial^\dagger\Lambda^2_{(p,q-1)}\;,
\label{gauge transf N=2}
\end{eqnarray}
and one can see again that gauge fixing the compensator to zero would restrict the gauge parameters to be transverse.

Let us turn now to the path integral quantization of the $U(2)$ spinning particle. If the particle is quantized on a circle with external gravity, one finds a representation for the QFT one-loop effective action of the $(p,q)$-form in a gravitational background. We will present the corresponding heat kernel expansion in terms of local Seeley-DeWitt coefficients.
In order to quantize the spinning particle, we have to take into account the gauge fixing of local worldline symmetries and Faddeev-Popov determinants. Because of the topology of the worldline circle, one is left with three modular integrals: the usual one over $\beta$, being the proper length of the circle, and two angular integration over $\theta$ and $\phi$, taking into account the gauge invariant Wilson loops of the two $U(1)$ factors.
The resulting partition function is given by
\begin{eqnarray}\nonumber
Z_{p,q}[g]&=&\int_0^\infty\frac{d\beta}{\beta}\int_0^{2\pi}\frac{d\theta}{2\pi}\int_0^{2\pi}\frac{d\phi}{2\pi}\,\mu(\phi,\theta)\times\\
&\times&\int\frac{d^dx_0d^d\bar x_0}{(2\pi\beta)^d}g(x_0,\bar x_0)\,\left\langle e^{-S_{\rm int}}\right\rangle\;,
\label{effective action defined}
\end{eqnarray}
where $x_0$ is an arbitrary fixed spacetime point, $S_{\rm int}$ is the interaction part of the spinning particle action and $\mu(\phi,\theta)$ is the modular measure given by
\begin{eqnarray}\nonumber
\mu(\phi,\theta)&=&e^{-i(p+1-d/2)\phi}e^{i(q+1-d/2)\theta}\times\\
&\times&\left(2\cos\frac{\phi}{2}\right)^{d-2}\left(2\cos\frac{\theta}{2}\right)^{d-2}\;.
\label{modular measure}
\end{eqnarray}
After performing the worldline perturbative computation, and evaluating the modular integrals\footnote{In the modular integration one encounters poles along the integration path. Detailed explanation of the prescription to deal with such poles can be found in \cite{Bastianelli:2012nh}} we organize the effective action expansion, up to order $\beta^2$, as follows:
\begin{eqnarray}\nonumber
&&\!\!\!\!\!\!\!\!\!\!\!\!\!\!\!\!Z_{p,q}[g]=\int_0^\infty\frac{d\beta}{\beta}\int\frac{d^dx_0d^d\bar x_0}{(2\pi\beta)^d}g(x_0,\bar x_0)\,v_1\left\{1+v_2\beta\,R\right.\\
&&\!\!\!\!\!\!\!\!\!\!\!\!\!\!\!\!\left.+\beta^2\left[v_3\left(R_{\mu\bar\nu\lambda\bar\sigma}\right)^2+v_4\left(R_{\mu\bar\nu}\right)^2
+v_5\,R^2+v_6\nabla^2R\right]\right\}\,.\nonumber\\
&&
\label{effective action coeff}
\end{eqnarray}
The Seeley-DeWitt coefficients $v_i$ are given by
\begin{eqnarray}\nonumber
&&\!\!\!\!\!\!\!\!\!\!\!\!v_1=\left(\frac{d-2}{p}\right)\left(\frac{d-2}{q}\right)\;,\quad v_2=\frac16-12k_1\;,\\ \nonumber
&&\!\!\!\!\!\!\!\!\!\!\!\!v_3=\frac{1}{180}-k_2+k_3\;,\quad v_4=-\frac{1}{360}-k_1+4k_2-2k_3\;,\\
&&\!\!\!\!\!\!\!\!\!\!\!\!v_5=\frac{1}{72}+k_1-3k_2+k_3\;,\quad v_6=\frac{1}{60}-k_1\;,
\label{SdW coeff}
\end{eqnarray}
where the numerical factors $k_i$ read
\begin{eqnarray}\nonumber
k_1&=&\frac{p(d-2-q)+q(d-2-p)}{24(d-2)^2}\;,\\ \nonumber
k_2&=&\frac{p(d-2-p)+q(d-2-q)}{24(d-2)(d-3)}\;,\\
k_3&=&\frac{p(d-2-p)q(d-2-q)}{2(d-2)^2(d-3)^2}\;.
\label{k coeff}
\end{eqnarray}
A few comments are now in order. First of all, let us stress that the overall coefficient $v_1$ gives the number of propagating degrees of freedom. For $p$ or $q$ greater than $d-2$ it vanishes and cannot be factored out in (\ref{effective action coeff}). Other coefficients can and indeed are non-vanishing, and represent the topological contribution of a non-propagating form. As a second remark we should notice that the result (\ref{k coeff}) holds only for $d>3$.
In $d=2$ only scalars propagate, and all the $k_i$ vanish. In $d=3$ instead, they have the form
\begin{equation}\label{k coeff d=3}
k_1=\frac{p(1-q)+q(1-p)}{24}\;,\; k_2=\frac{p+q}{24}\;,\; k_3=\frac{pq}{2}
\end{equation}
Finally, we stress that the $k_i$ coefficients in (\ref{k coeff}) were used to make manifest the symmetry under the exchanges $p\leftrightarrow q$ and $p\leftrightarrow d-2-q$. The first exchange is related to the symmetry under complex conjugation, that states the equivalence $Z_{p,q}[g]=Z_{q,p}[g]$ and is exact. The second exchange relates forms with Hodge dual curvatures and is more subtle. Despite the manifest symmetry in the coefficients (\ref{k coeff}), it is not an exact symmetry at the quantum level, and one can already see that in $d=3$ the $k_i$ (\ref{k coeff d=3}) are not invariant under $p\leftrightarrow 1-q$. In general, for higher dimension $d$ the mismatch appears in higher order Seeley-DeWitt coefficients.
The spinning particle model, however, allows us to find an exact non-perturbative result for the mismatch, that is purely topological. The derivation can be found in \cite{Bastianelli:2012nh}, and one has\footnote{We denoted $Z_{p,q}=\int_0^\infty\frac{d\beta}{\beta}\mathcal{Z}_{p,q}(\beta)$.}
\begin{eqnarray}\nonumber
&&\!\!\!\!\!\!\!\!\!\mathcal{Z}_{d-2-q,d-2-p}(\beta)-\mathcal{Z}_{p,q}(\beta)=(-)^{q+d}\mathcal{Z}^{\rm top}_{d-1,q}(\beta)\\ \nonumber
&&\!\!\!\!\!\!\!\!\!+(-)^{p+d}\mathcal{Z}^{\rm top}_{p,d-1}(\beta)+(-)^{p+q}\mathcal{Z}^{\rm top}_{d-1,d-1}(\beta)\\ \nonumber
&&\!\!\!\!\!\!\!\!\!+(d-1-p)(-)^{p+q}\sum_{m=0}^q(-)^m(q+1-m)\,{\rm ind}(\Omega^{m,0},\bar\partial)\\ \nonumber
&&\!\!\!\!\!\!\!\!\!+(d-1-q)(-)^{p+q}\sum_{n=0}^p(-)^n(p+1-n)\,{\rm ind}(\Omega^{n,0},\bar\partial)\\
&&\!\!\!\!\!\!\!\!\!+(-)^{p+q}\left[\left(p+1-\frac d2\right)\left(q+1-\frac d2\right)-\frac{d^2}{4}\right]\,\chi(\mathcal{M})\;.\nonumber\\
&&
\label{mismatch}
\end{eqnarray}
In the above formula ${\rm ind}(\Omega^{m,0},\bar\partial)$ is the Dolbeault index twisted by the $(m,0)$-form bundle, whose expression can be found in \cite{Bastianelli:2012nh,Nakahara}. $\chi(\mathcal{M})$ is the Euler characteristics of the manifold, and the top forms $Z_{p,d-1}$, $Z_{d-1,q}$, $Z_{d-1,d-1}$ are related to the Ray-Singer analytic torsion \cite{Ray:1973sb} via
\begin{equation}
Z_{p,d-1}^{\rm top}=2\sum_{n=0}^p(-)^{n+1}(n+1)\ln T_{d-p+n}(\mathcal{M})\;.
\end{equation}
Having presented the expansion for the effective actions up to order $\beta^2$, it is possible to check the mismatch (\ref{mismatch}) only in $d=2$ and $d=3$. These cases have been considered in \cite{Bastianelli:2012nh} and agree with (\ref{mismatch}), giving a nontrivial check.

\section{Conclusion}
In this short contribution, we presented the $U(N)$ spinning particles, introduced in \cite{Marcus:1994mm} and subsequently studied in \cite{Bastianelli:2009vj}, where they were shown to describe a class of gauge invariant higher spin equations, close in form to Fronsdal-Labastida equations for real spacetimes and to Maxwell-like equations recently introduced in \cite{Campoleoni:2012th}. We briefly described the possibility of defining the model on a K\"ahler manifold, and finally focused on the $U(2)$ model on an arbitrary K\"ahler space. We presented field equations for differential $(p,q)$-forms as well as the heat kernel expansion of their effective actions and issues related to Hodge duality. Even if a genuine spacetime interpretation is somehow prevented by the complex target space, these models can provide interesting insights in the general problem of higher spin field theories, sharing crucial properties such as the appearance of constrained gauge invariance and compensator fields \cite{Francia:2002aa}. When quantized on K\"ahler backgrounds, they show quite a rich structure, such as the topological issues related to Hodge duality, or the coupling to the $U(1)$ part of the  K\"ahler holonomy, and could be a useful instrument in K\"ahler geometry.

\section*{Acknowledgement}
I would like to thank Fiorenzo Bastianelli and Carlo Iazeolla for working with me on this topic, and Dario Francia for valuable discussions.

\end{multicols}
\end{document}